\documentstyle[prl,twocolumn,aps]{revtex}
\draft 
\begin{document}
\title{
Asymptotic tunneling conductance in Luttinger liquids}
\author{F. Guinea \\}
\address{
	Instituto de Ciencia de Materiales. CSIC. \\
	Cantoblanco. E-28049 Madrid. Spain.}
\author{G. G\'omez-Santos \\}
\address{
	Departamento de F{\'\i}sica de la Materia Condensada. \\
	Universidad Aut\'onoma. E-28049 Madrid. Spain.}
\author{M. Sassetti \\}
\address{
        Istituto di Fisica di Ingegneria, INFM. \\
        Universit\'a di Genova. I-16146 Genova. Italy.}
\author{M. Ueda \\}
\address{
	Department of Physical Electronics, Hiroshima University, \\
	Higashi-Hiroshima 724, Japan.}
\maketitle
\date{\today}
\begin{abstract}
Conductance through weak constrictions in Luttinger liquids is shown to vanish
with frequency $\omega$ as 
$c_1 \omega^2 + c_2 \omega^{2/g - 2}$, where $g$ is a
dimensionless parameter characterizing the Luttinger liquid phase, and 
$c_1$ and $c_2$ are nonuniversal constants.
The first term arises from the ^^ Coulomb blockade' effect and
dominates for $g < 1/2$, whereas the second results from
eliminating high-energy modes and dominates for $g > 1/2$. 
\end{abstract}
\vskip 0.2cm
\pacs{PACS numbers: 75.10.Jm, 75.10.Lp, 75.30.Ds.}

\narrowtext

	Tunneling in Luttinger liquids has 
lately attracted a great deal of attention 
because this has been suggested to be the best way 
to characterize the Luttinger liquid phase and because 
a simple example of Luttinger liquids is nowadays accessible 
experimentally, the edge current in the fractional quantum Hall 
effect~\cite{experiments}.

	While a bulk Luttinger liquid is well understood~\cite{Haldane},
the tunneling problem has only recently been initiated by Kane and 
Fisher~\cite{KF}, 
who mapped the problem onto the Schmid model~\cite{Schmid}, which has 
extensively been studied in connection with Josephson junctions. 
They showed that the duality exhibited by this model is also applicable 
to the Luttinger problem. Then using a well-established renormalization-group 
(RG) approach~\cite{RG1,RG2}, they suggested that for the weak-barrier and
repulsive case, the low-temperature conductivity of a junction 
between Luttinger liquids should vanish as $T^{2/g-2}$,
where $g$ is a dimensionless parameter which depends on
the strength of the interactions (see below).

	It is thus surprising that Mak and  Egger~\cite{Mak}, 
using the same model and mapping, suggested 
the $T^2$ law in the case of the weak-barrier and repulsive case
on evidence from real-time Montecarlo simulations. 
In this case a finite cut-off is employed to extract the low temperature
regime. Note that expansions around $g = 1/2$, in which the
cutoff is set to infinity give different results~\cite{weiss}. 
Other Montecarlo work in imaginary-time, however, 
reported conflicting results~\cite{Montecarlo}.

	In view of current experimental and 
theoretical efforts to understand the
transport properties of Luttinger liquids, it seems urgent to clarify the 
situation. In this paper, we show that the asymptotic
low frequency conductance,
in general, obeys 
$c_1 \omega^2 + c_2 \omega^{2/g-2}$, and that depending
on whether $g < 1/2$ or $g > 1/2$, one term dominates or the other.

	Consider first the weak-barrier, attractive case.
The Hamiltonian that describes a free bosonic field in a local periodic 
potential is given by~\cite{KF,Mak}
\begin{equation}
	{\cal H} = {1\over 2} \int dx \left( 
	g \Pi ( x )^2 +  g^{-1} (\partial_x \phi
	( x ) )^2 \right) + V \cos \left( \sqrt{4 \pi } \phi ( 0 ) \right),
\end{equation}
where the harmonic field describes the dynamics of a perfect 
Luttinger liquid, and the periodic term describes the effect of
the barrier.
The dimensionless parameter $g$ characterizes the decay of the electronic
Green functions and depends on the nature of the interactions: 
$g < 1$ corresponds to repulsive interactions, and $g > 1$ to attractive ones.
As shown in Ref.~\cite{KF}, the bosonic 
degrees of freedom other than $\phi ( 0 )$ can be integrated out,
and a simple scaling equation for $\tilde{V} = V / \omega_c$ 
($\omega_c$ is the cutoff) follows:
\begin{equation}
	{{\partial \tilde{V}}\over{\partial l}} =-
	\left( 1 - \frac{1}{g} \right) \tilde{V},
\end{equation}
where $l = - \log ( \omega_c )$. 
Thus $\tilde{V}$ is an irrelevant operator for attractive
interactions and weak barriers, when the problem scales
towards a free fixed point, and the model is solvable in the low-energy limit. 
In addition, the weak-barrier, 
attractive case can be mapped onto the strong-barrier, 
repulsive case, and vice versa. The low energy properties of the junction are
determined by the scaling of $\lim_{\omega_c \rightarrow 0} 
\tilde{V}  \propto
{\omega_c}^{1 - 1 / g}$ 
near the fixed point,
leading to a conductance for the strong barrier repulsive case $(g<1)$:

\begin{equation}
	G \sim \omega^{\frac{2}{g} - 2} \sim T^{\frac{2}{g} - 2}
\end{equation}

	Consider next the weak-barrier, repulsive case.
The fact that the value of $g$ is not renormalized suggests a simple
interpolation~\cite{RG1,RG2,KF} between the strong-barrier and weak-barrier
regimes.
As $\tilde{V}$ for repulsive interactions 	
and weak barriers grows upon scaling, 
we expect to eventual crossover to the case of 
strong barriers, so the low-energy properties
of the system are described by a scaling such as eq.~2,equations like eqs.~(3).
As analyzed in detail below, however, this conjecture overlooks
excitations that can modify the dependence shown in eqs.~(3).

	Since the periodic potential becomes relevant for the repulsive case, 
we need to infer the nature of the fixed point the Hamiltonian flows to.
We assume that this fixed point is well described by the self-consistent 
harmonic approximation (SCHA) to this model~\cite{SCHA}, where the
trial Hamiltonian reads
\begin{equation}
	{\cal H}_{\rm SCHA} = {1 \over 2}
	\int dx \left( g \Pi ( x )^2 + 
	g^{-1} ( \partial_x \phi ( x ) )^2 \right)
	+ {1 \over 2} 4 \sqrt{\pi} V'\phi ( 0 )^2.
\end{equation}
	The strength of this potential is then given by
\begin{equation}
	V' = V \left( {V\over{\omega_c}} \right)^{{g\over{1 - g}}}.
\end{equation}

	There are various reasons for 	
this approximation being essentially correct:

	- We know from the RG analysis that 
the system flows towards a localized regime.
Any scheme that reproduces this effect,
and the associated zero-point fluctuations, should be
a good approximation. 

	- The strength of the harmonic potential $V'$ coincides
with the effective value of $V$ at the scale when $\tilde{V} \sim 1$.
At this scale the flow equation (2) ceases to be valid. Thus the
approximation contains, basically, the same information as the
RG approach.

	- If the crossover from the weak coupling to the strong coupling
regimes is described by a single energy scale,
this scale should be given, approximately, by eq. (5).
This assumption correctly describes the low-energy dynamics of 
the related model of a two level system interacting ohmically
with a dissipative bath~\cite{TLS1}. The low-energy behavior
of the Kondo model can also be expressed in this way~\cite{Kondo}.

	In calculating the conductivity of the model,
it is illuminating to analyze the new contribution to the Hamiltonian
in terms of fermion operators. Using the relation for the
fermion density $\rho ( x ) = \partial_x \phi/\sqrt{\pi}$, we can write
\begin{equation}
	\phi ( 0 )^2 =\pi( Q_R - Q_L )^2/e^2,
\end{equation}

where $Q_{L,R}$ are the total charges on the right and left
of the junction. This new term represents a charge-charge interaction,
studied extensively in connection with the problem of Coulomb 
blockade~\cite{CB1,CB2}.
To understand the origin of this term, 
we should note that the initial backward scattering term can be expressed as
$\sim V \exp ( 2 \pi i( Q_R - Q_L )/e )  + {\rm h.c.}$ 

This term
suppresses small fluctuations of the charge across the junction.        

	For an attractive Luttinger model in the 
strong-barrier limit, the scaling increases the
effective dimensionless hopping. Thus we expect the low-energy properties
to be determined by a Hamiltonian similar to eq. (5). The low-energy 
modes  correspond, in this case, to small fluctuations in the relative
phase of the ends of the two semi-infinite systems. This situation is 
opposite to the one considered earlier. Upon scaling, the barriers
become weaker and eventually the two chains can no longer be separated. 
Phase slips across the barrier are not allowed. The junction resembles
a classical Josephson junction, and the harmonic potential in eq. (4)
gives the plasma frequency. The bulk system, however, is gapless,
because of the dimensionality. Some of the bulk low-energy excitations
hybridize with the plasma mode, modifying the properties
of the junction. 

	For the case of weak barriers and repulsive interaction,
calculation of the conductance within the
framework discussed above is straightforward. The effect of
an applied voltage can be described by a term in the Hamiltonian~\cite{KF}
\begin{equation}
	{\cal H}_V = V_{\rm appl} ( Q_L - Q_R ) \sim V_{\rm appl} \phi ( 0 ),
\end{equation}
where $V_{\rm appl}$ is the applied voltage. The conductance 
is given by
\begin{equation}
	G ( \omega ) = {|{\langle 0 | j | \omega \rangle |^2}\over{\omega}},
\end{equation}
where $j = e\dot{ \phi } ( 0 )/\sqrt{\pi}  $, 

so that
$ \langle 0 | j | \omega \rangle \sim \omega \langle 
0 | \phi ( 0 ) | \omega \rangle$.
The latter quantity can be obtained from the decomposition of 
$\phi ( 0 )$ in normal modes. The square of the amplitude of
a mode of frequency $\omega$ at the position of the barrier
is proportional to the transmission coefficient of the barrier, $\tilde{T}$.
A simple calculation gives
\begin{equation}
	\tilde{T} \sim {{\omega^2}\over{{V'}^2}},
\end{equation}
which means that the barrier is perfectly reflecting at
zero energy, and that the conductance scales as
\begin{equation}
	G ( \omega ) \sim \tilde{T} \sim \omega^2
\end{equation}

The appearance of
the $\omega^2$ law has been extensively discussed in connection 
with junctions that exhibit Coulomb blockade\cite{CB2}.
It is due to the small amplitude charge fluctuations across
the junction. Alternatively, ${\cal H}_{SCHA}$ describes
the damped quantum oscillator. We can use the standard analysis
of this problem~\cite{Weiss}, we can write:

\begin{equation}
	G ( \omega ) = \frac{e^2 g}{2 \pi} \frac{\omega^2}
	{{2 \pi V'}^2 + g^{-2} \omega^2}
\end{equation}  

	These results can be extended to finite temperatures. 
A straightforward calculation (for $t \gg \hbar {V'}^{-1}$) gives:

\begin{eqnarray}
	2\langle \dot{ \phi } ( t )  \dot{ \phi } ( 0 ) \rangle &=
	&\langle \{ \dot{ \phi } ( t ) , \dot{ \phi } ( 0 )  \} 
	\rangle +
	\langle [ \dot{ \phi } ( t )  , 
	\dot{ \phi } ( 0 )  ] \rangle \nonumber \\
	 &\propto &\int_0^{\infty} d \omega 
	\frac{\omega^3}{1 - e^{ - \beta \omega}} \left(
	e^{i \omega t} + e^{- \beta \omega} e^{- i \omega t} \right)\nonumber\\
	 &= &\frac{2 \pi^4 T^4 [ 1 + 2 \cosh^2 ( \pi t T ) ]}
	{ \sinh^4 ( \pi t T )} + i \delta ''' ( t )
\end{eqnarray}

	where the commutator  and anticommutator 
in this expression are related to the noise and dissipation in thermal 
equilibrium~\cite{KF2}.
As the hamiltonian is harmonic, the imaginary part is independent of
temperature. 

	The previous analysis includes only the small-amplitude 
charge oscillations near the minima of the potential. We can also
include discrete charge transfer processes by analyzing transitions
between neighboring minima (see figure 1). The wavefunction corresponding
to a minimum centered around $\phi = \sqrt{ \pi } n  $, $| 2 \pi
n \rangle$, is given by
the ground state of the hamiltonian:

\begin{equation}
	{\cal H} = {\cal H}_{\rm SCHA} -  2 \pi n V' \phi ( 0 ).
\end{equation}

	The hopping amplitude between neighboring minima
can be calculated using the wavefunctions obtained from
eq. (13) as a variational basis for the original 
hamiltonian, eq. (1). A correction to the ground state energy
of the form $2 t_{eff} \cos ( q )$  is obtained, where $q$ is
the band index, and: 

\begin{equation}
	t_{eff} = \langle 0 | {\cal H} | 2 \pi \rangle
	- \langle 0 | {\cal H} | 0 \rangle \langle 0 | 2 \pi \rangle =
	V' \langle 0 | 2 \pi \rangle \left( 1 - \frac{\pi^2}{4} \right) 
\end{equation} 

	${\cal H}$ is given in eq. (1).
The value of the overlap, $\langle 0 | 2 \pi \rangle$, 
can be computed by 
noting that eqs. (4) and (13) are
related by a canonical transformation.
At low energies, and decomposing $\phi ( 0 )$ in the normal
modes of ${\cal H}_{SCHA}$ (eq. 4 ), the required transformation is:

\begin{equation}
	U b_k^+ U^{-1} = b_k^+ + \pi \sqrt{\frac{2}{g k}}.
\end{equation}

	Note that $U$ does not depend on $V'$. 

	From this analysis, we can infer the contribution of the
interminima hopping to the junction conductivity:

\begin{equation}
	\hat{j} ' \propto i e V' ( U - U^{-1} )
\end{equation}

	and:

\begin{equation}
	\langle \hat{j} ' ( t ) \hat{j} '  ( 0 ) \rangle \sim
	( V' t )^{2 / g}
\end{equation}

	This result implies that there is a contribution to the
conductance which scales as $G ' \sim \omega^{2/g - 2}$.
Extending the result to finite temperatures, we find a conductance
which scales as: $G' \sim T^{2/g - 2}$. Thus, the inclusion
of interminima processes leads us to the expressions in
the conductance derived in~\cite{KF}. Alternatively, we can also
say that this scaling behavior arises when the periodicity
of the variable $\phi$ is restored. Note that, in terms of the
backscattering potential $V$ (eq. 1), the correlation function
(16) depends only on the combination $V^{1/(1-g)} t$. This confirms
the existence of a universal function which interpolates
from large to small scales~\cite{KF}.

	We can also study higher order processes induced 
by the residual interminima couplings. An expansion in terms of
the hopping is equivalent to the expressions obtained in
the strongly localized, tight binding formulation of the
problem. An elegant scheme developed for the treatment of
noise in this limit can be found in~\cite{noise}.
The study reported in~\cite{noise} can also be used to 
compute the conductance to any order of the interminima
hopping amplitude. To second order, we obtain 
the $G \propto \omega^{2/g - 2}$ behavior, already 
discussed ( and calculated for the first time in~\cite{KF}).
It is interesting to note that, to fourth order, 
keeping a finite cut-off
we obtain
another contribution of the type $G \propto \omega^2$.

	This contribution is of ^^ ^^ interdipole " origin. At finite
temperatures, scaling arguments imply that
the same processes lead to a $G \propto T^2$
dependence. The diagrams which lead to this effect were already
discussed in connexion with Coulomb blockade~\cite{CB1,CB2},
in the same tight binding limit.

	 Note that the 
$G \propto \omega^2$ dependence in the SCHA approximation
does not give rise to a finite conductance at finite temperature,
due to the confinement of $\phi$ induced by the parabolic potential.
The interdipole processes which appear when the periodicity of
$\phi$ is restored, on the other hand, do translate into a
finite conductance at finite temperatures. In both cases, however,
we are dealing with low amplitude charge fluctuations, in which
the total charge transferred is less than the charge of one electron.

	While it seems obvious that low amplitude polarization fluctuations
contribute to the frequency dependence of the conductance, their
role in the d. c. conductance at finite temperatures is,
at first sight, somewhat unphysical. Their influence on the 
d. c. current requires a detailed analysis of the junction plus
the external circuit. Usually,
the current flowing through the junction
is computed from matrix elements like $\langle Q | {\cal H}_{tunnel}
| Q + e \rangle$~\cite{AL}. This scheme implicitly assumes
that the external part of the circuit, the battery, can only 
gain or lose charge in discrete units. This needs not be the case.
The external battery keeps the temperature, and the chemical
potential fixed. Hence, its typical response times are
$\hbar \tau^{-1}_{batt} \sim \hbox{max} ( T , eV )$. 
In that time, the charge which moves across the junction is
$G \tau_{batt} V$. Thus, if $G \ll \frac{e^2}{\hbar}
\hbox{max} ( 1 , \frac{T}{eV} )$,
the battery should be able to take charge in arbitrary 
units~\cite{quantum}. 

	It is illustrative to describe the two contributions to
the conductivity in statistical mechanics
terms. The nonanalytic term, $\omega^{2 / g - 2}$, is due to topological
excitations, while the analytic part, $\omega^2$, comes
from low amplitude spin waves~\cite{instantons}. 
The latter dominates for $g < 1/2$.

	In conclusion, we present here an analysis of the strong
coupling fixed point to which the tunneling hamiltonian, eq. (1),
flows at low energies. The behavior of the system is shown to be
closely related to that found in junctions which exhibit Coulomb 
blockade. We have discussed in detail the possible contributions
of low energy charge fluctuations to the frequency and temperature
dependence of the conductance. These processes, which involve small
charge transfers across the junction, can be observed in a d. c.
experiment at finite temperatures, {\it provided that the charge at
the junction cannot be considered quantized}.  
	 
	We are thankful to U. Weiss, for introducing us to this problem,
and for many interesting discussions. This work has been supported, in
part, by CICyT, Spain (grant MAT91-0905).

\figure{Figure 1.
Sketch of the parabolic potential defined in the self consistent
harmonic aproximation. The arrow indicates the interminima processes
which need to be considered separately (see text).}

\begin{references}
\bibitem{CB2}
M. Ueda and F. Guinea, Z. Phys. B {\bf 85},
413  (1991) (note that in equation (5) of this reference,
the exponent $2 + g$ should be replaced by 2).  
F. Guinea and M. Ueda, Phys. Rev. B {\bf 49}, 5722 (1994).
\end{references}
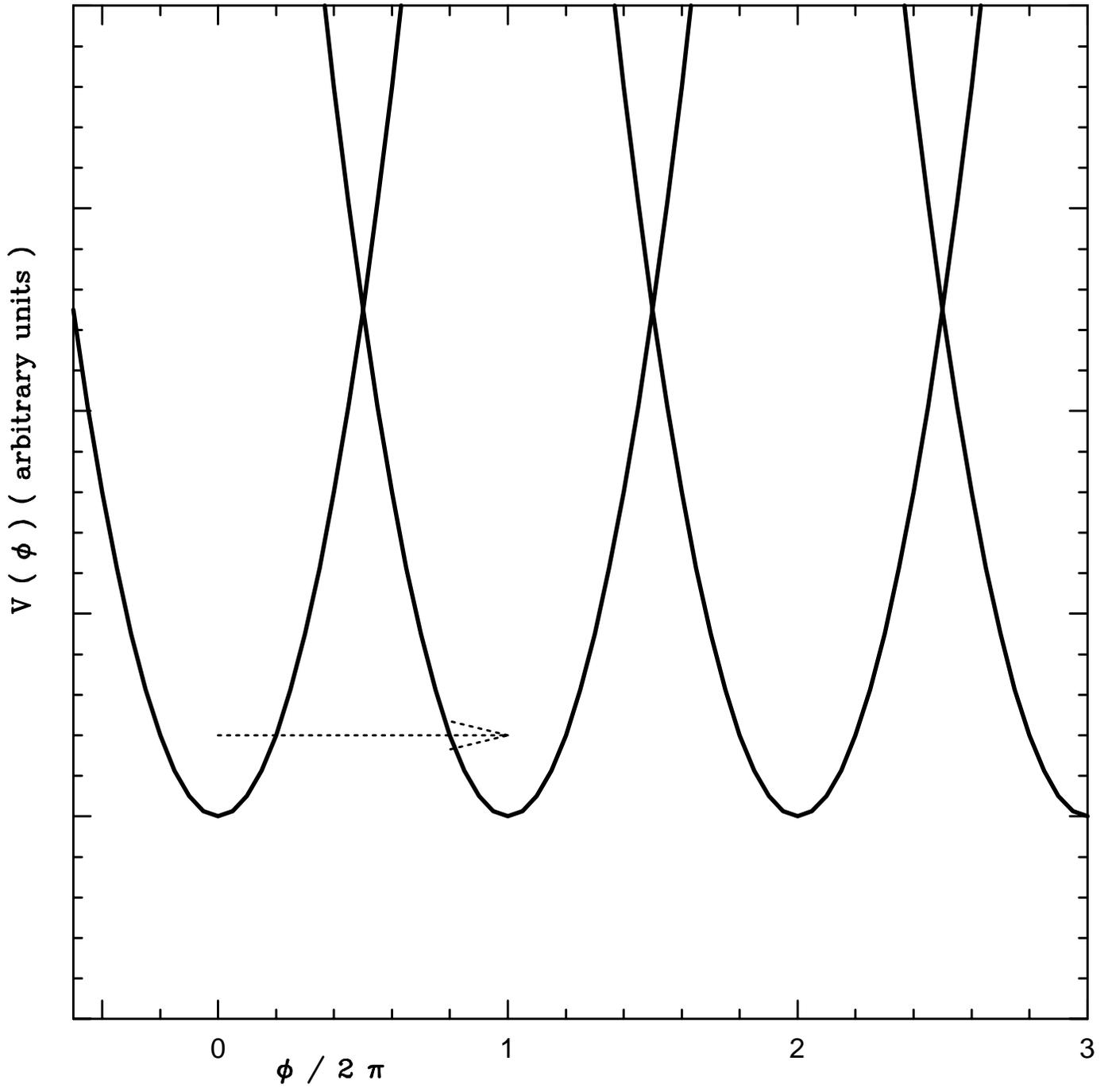
\end{document}